\documentstyle[12pt]{article}

\begin{document}

\centerline{\Large A surface theoretic model of quantum gravity}
\vspace{0.5cm}
\centerline{\rm Junichi Iwasaki}
\vspace{0.2cm}
\centerline{\rm Department of Physics,
                Universidad Autonoma Metropolitana, Iztapalapa}
\centerline{\rm AP 55-534, Mexico DF 09340, Mexico}
\vspace{0.2cm}
\centerline{\rm E-mail:\ iwasaki@abaco.uam.mx}
\vspace{0.2cm}
\centerline{March 29, 1999; revised on November 30, 1999}


\begin{abstract}
A surface theoretic view of non-perturbative quantum gravity as 
``spin-foams" was discussed by Baez.
A possibility of constructing  such a model was studied
some time ago based on (2+1) dimensional general relativity as 
a reformulation of the Ponzano-Regge model in Riemannian spacetime.
In the present work, a model based on (3+1) dimensional general relativity 
in Riemannian spacetime is presented.
The construction is explicit and calculable in details.
{}For a physical application, 
a computation formula for
spacetime volume density correlations is presented.
Remarks for further investigations are made.
\end{abstract}


\section{Introduction}\label{sec:introduction}

A surface theoretic view of non-perturbative quantum gravity as 
``spin-foams" was discussed by Baez \cite{JB}.
A possibility of constructing  such a model was studied
some time ago based on (2+1) dimensional general relativity as 
a reformulation of the Ponzano-Regge model in Riemannian spacetime 
\cite{JI}.
In the present work, a model based on (3+1) dimensional general relativity 
in Riemannian spacetime is presented.
The main difference is that (3+1) general relativity is not a topological
field theory but has local degrees of freedom while (2+1) general relativity
is identical to the non-degenerate sector of
BF theory (a topological field theory) 
in (2+1) dimensions.
The construction is explicit and calculable in details.
This model allows one to investigate the difference between 
the quantum theories for general relativity and BF theory  
from a surface theoretic point of view.
{}For a physical application, a computation formula for
spacetime volume density correlations is presented.
{}For a family of models allowing spin-foam interpretation of 
quantum gravity, see the references \cite{JB,JI,MR,CR,BC,MS}.
In particular, surface theoretic views are discussed by
Baez \cite{JB}, Reisenberger \cite{MR},
Reisenberger and Rovelli \cite{CR} and the author \cite{JI}.\footnote
{The kind of surfaces playing important roles is called
in terms of different names by different authors. 
The name ``spin-foam'' is due to Baez.}
In the present work,
the calculation technologies developed in the connection representation
\cite{AL} and the loop representation \cite{DR} of
non-perturbative canonical quantum gravity play important roles.

Among the models revealing aspects of the spin-foam interpretation, 
the philosophy of the present model is similar to that of 
the Reisenberger model \cite{MR} in the sense that both models are based on
a path integral of general relativity.
The other models are based on lattice topological field theories 
\cite{JI,BC},
canonical quantization of general relativity \cite{CR} or others 
\cite{MS}.
However, the technical difference of the present model from 
that of the Reisenberger model is
that the present model uses multiple pairs of lattices.
In each pair, one lattice is dual to the other.
The pairs of lattices are used to regularize the exterior product of
two forms.
The exterior product of two 2-forms, which is present in the action
functional of general relativity in 4-dimensions, 
is naturally defined on the lattice
at the intersection of a face and its dual face.
In addition, the degenerate and non-degenerate sectors of 
the theory are clearly characterized by the number of
pairs of lattices.

In a limit, the partition function of the model is reduced to 
a product of copies of the partition function of lattice BF theory, 
each of which is independently defined on one of the lattices.\footnote
{The partition function of lattice BF theory in 4-dimensions
defined on a single lattice was written by Ooguri \cite{HO}.}
This fact implies that 
the contribution from each lattice is decoupled in this limit and
the use of the multiple lattices for   
topological field theory is redundant. 
In general, however, 
the contribution from each lattice cannot be decoupled and
the partition function of the model is different 
from that of lattice BF theory.
This difference is supposed to correspond to the fact that
general relativity contains local degrees of freedom while 
BF theory does not.
In this way, the model provides a way of clarifying local degrees of freedom
of quantum gravity.

The present model is formulated such a way that the inclusion of 
local degrees of freedom of general relativity, 
which is absent in BF theory,
is incorporated to coefficients in the partition function
and the ``trivialization" of the coefficients leads to 
(a product of copies of) lattice BF theory.
The computation of the coefficients is done by the regularization
in terms of the multiple pairs of lattices.
Each pair consists of two lattices dual to each other.
The ``trivialization'' corresponds to the limit mentioned above.

In the next section, the partition function of the model is presented 
at the beginning, and then the rest of the section is devoted to
the determination of coefficients in the partition function, followed
by the presentation of a computation formula for spacetime volume density
correlations.
Remarks for further investigations are made in the conclusion section.


\section{The model}\label{sec:model}

The partition function of the model is
\begin{eqnarray}
&& Z_{\beta,\lambda}[\xi]:=
\prod_{p=1}^{P}\prod_{e\in\Delta_p\Delta_p^*}\int dU(e)\times
\prod_{f\in\Delta_p\Delta_p^*}\sum_{j}C_{\beta,\lambda}^{(j)}[\xi]
\chi_j(\prod_{e\in f}^{path}U(e)),
\label{eq:partition}
\\&&
C_{\beta,\lambda}^{(j)}[\xi]:=
\sum_{\{l\}}W_{\beta,\lambda}(\{l\},\xi)\Omega^{(j)}(\{l\}).
\label{eq:coefficients}
\end{eqnarray}
Here, $\beta$ is $1$ or $i$ (the imaginary unit number) and $\lambda$ is
a real or imaginary number.
$\xi$ is an external field coupled to spacetime volume density, 
$\int d^4x \sqrt{g}\xi(x)$.
This field is a mathematical tool to calculate spacetime volume density
correlations.
$P$ is the number of pairs of lattices and
$\Delta_p$ and $\Delta_p^*$ are a pair of lattices dual to each other.
The index $p$ runs from $1$ through $P$. 
A family of spin-foams are defined in terms of each of the lattices.
More details of the lattices and spin-foams are described below.
$\chi_j(U)$ is the character of SU(2) element $U$ in the spin-j 
representation.
$\prod^{path}$ means the path ordered product.
$dU$ is the Haar mesure for SU(2) element $U$.
$e\in\Delta_p$ and $f\in\Delta_p$ mean that they are an edge (1-cell) and 
a face (2-cell) of $\Delta_p$ respectively.
$e\in f$ means that $e$ is an edge shared by $f$ and other faces;
in other words, $e$ is a boundary of $f$.
$j$ and $l$ are spins associated with the faces 
taking values 0, $1\over2$, 1, ${3\over2}\cdots$.
$W_{\beta,\lambda}(\{l\},\xi)$ and $\Omega^{(j)}(\{l\})$ are
coefficients to be determined below.
A particular property of these coefficients are as follows.
When $\beta=1$, $\lambda=0$ and $\xi=0$, 
then $C_{1,0}^{(j)}[0]=\prod_{p=1}^{P}\prod_{f\in\Delta_p\Delta_p^*}(2j+1)$
with $j$ associated with each $f$.
Accordingly the contributions from the different lattices are decoupled
and the partition function becomes
\begin{eqnarray}
&&Z_{1,0}[0]=
\prod_{p=1}^{P}
\prod_{e\in\Delta_p\Delta_p^*}\int dU(e)\times
\prod_{f\in\Delta_p\Delta_p^*}\sum_{j}(2j+1)
\chi_{j}(\prod_{e\in f}^{path}U(e))
\nonumber\\&&=
\left[
\prod_{e\in\Delta}\int dU(e)\times
\prod_{f\in\Delta}\sum_{j}(2j+1)
\chi_{j}(\prod_{e\in f}^{path}U(e))\right]^{2P}.
\end{eqnarray}
Notice that this is not just the partition function of lattice SU(2) 
BF theory in (3+1) dimensions, a topological invariant, but
the multiple power of it.
We have used the fact that the partition function of lattice BF theory
does not depend on the choice of lattice and let $\Delta$ represent
one of the lattices.
A line of studies of lattice BF theory is detailed in the reference
\cite{BC}.


The lattices $\Delta_p$ and $\Delta_p^*$ used in the partition function are
defined as follows.
They are 4-dimensional piecewise linear cell manifolds consisting of
0-cells (vertices),
1-cells (edges), 2-cells (polygons), 3-cells (polyhedrons) and 4-cells.
They are dual to each other.
That is,
every k-cell of $\Delta_p$ (and $\Delta_p^*$) intersects with a (4-k)-cell of
$\Delta_p^*$ (and $\Delta_p$ respectively) at a point inside the cells.
We call the intersection of a k-cell of $\Delta_p$ and a (4-k)-cell 
of $\Delta_p^*$ the ``center" of each of the cells.
Every cell of $\Delta_p$ and $\Delta_p^*$ has one and only one center.
We define the two lattices such that the center of a k-cell is not on 
the boundary of but inside the cell.
The exception is that the center of a 0-cell is identified to itself.
A spin-foam is defined as a 2-dimensional sub-complex of $\Delta_p$ or
$\Delta_p^*$ with 2-cells labeled by spins and 1-cells labled by intertwiners.
A 2-cell labeled by the 0-spin is understood as the absence of the cell 
in the 2-dimensional sub-complex under consideration.

In the model, we use the multiple of such pairs of lattices
and denote the set of the lattices by $\{\Delta\Delta^*\}$ meaning
$\{\Delta_p,\Delta_p^*(p=1,\cdots, P)\}$.
In addition, we impose two conditions to the lattices. 
One condition is that
the position of the center of every face of $\Delta_p$ coincides with
the position of the center of a face of $\Delta_q$
for all $p,q=1,\cdots,P$.
This condition is needed in order to characterize the degenerate and 
non-degenerate sectors of the theory.
The other condition is that
$\Delta_p$ and $\Delta_p^*$ are isomorphic to $\Delta_q$ and $\Delta_q^*$
respectively for all $p,q=1,\cdots, P$ 
and hence the faces of them correspond one-to-one.
In other words, $\Delta_1,\cdots,\Delta_P$ are copies of 
the same (abstract) lattice
embedded in different locations of the spacetime manifold 
with a restriction due to the first condition and
so are their respective dual lattices  $\Delta_1^*,\cdots,\Delta_P^*$.
The positions of the one-to-one corresponding faces
of $\Delta_p$ and $\Delta_q$ under the isomorphism do not necessarily
coincide with each other.
The second condition regularizes the kind of lattices utilized for the model
and guaratees that copies of lattice BF theory independently defined
on single lattices produce an identical partition function.
(A way of constructing the pairs of lattices is discussed 
in Appendix \ref{app:pairs}.)


In order to determine the coefficients $W_{\beta,\lambda}(\{l\},\xi)$ and
$\Omega^{(j)}(\{l\})$ we compute a path integral of general relativity.
The action for general relativity we consider is the Plebanski action 
\cite{JP}.
It consists of two terms, 
both of which is spacetime diffeomorphism 
invariant separately in addition to their internal gauge invariance, and is
\begin{eqnarray}
&&
S[A,B,\phi]:=\int d^4x\delta_{ij}\tilde\epsilon^{abcd}B_{ab}^iF_{cd}^j
-{\lambda\over2}\int d^4x\phi_{ij}\tilde\epsilon^{abcd}B_{ab}^iB_{cd}^j.
\label{eq:action}
\end{eqnarray}
Here $a,b,c\cdots$ are spacetime indices $\{0,1,2,3\}$ and
$i,j\cdots$ are internal SU(2) adjoint indices $\{1,2,3\}$.
$\tilde\epsilon^{abcd}$ is the alternating tensor with density weight $1$.
$F_{ab}^i:=\partial_aA_b^i-\partial_bA_a^i+\epsilon^i_{\ jk}A_a^jA_b^k$
are the curvature of an SU(2) gauge connection $A_a^i$,
$B_{ab}^i$ is an SU(2) algebra valued 2-form field and $\phi_{ij}$ 
are scalar fields 
(traceless symmetric with respect to the SU(2) indices).
The dimensions of $A_a^i$, $B_{ab}^i$ and $\phi_{ij}$ are of length inverse,
length inverse squared and length inverse squared respectively.
$\lambda$ is a parameter with the dimension of length squared
and its presence can be understood
if one rescales $A_a^i$ and $B_{ab}^i$ appropriately.
The Euclidean sector is obtained by real valued fields and imaginary 
time coordinate.
The five scalar fields $\phi_{ij}$ are Lagrange multipliers forcing 
the non-trace parts of
$\tilde\epsilon^{abcd}B_{ab}^iB_{cd}^j$ 
(about the internal indices $i$ and $j$) 
vanish.
This condition implies the exsistence of non-degenerate tetrad fields
$e_a^0$ and $e_a^i$ such that 
$B_{ab}^i=e_a^{[0}e_b^{i]}+{1\over2}\epsilon^{i}_{\ jk}e_a^je_b^k$
if $\delta_{ij}\tilde\epsilon^{abcd}B_{ab}^iB_{cd}^j\ne 0$.
In other words, in its non-degenerate sector the action has
the value of the self dual Hilbert-Palatini action evaluated on the tetrad.
The first term of (\ref{eq:action}) is known to be the action of 
the BF theory, a topological field theory.
The second term amounts to introduce local degrees of freedom contained
in general relativity.


We define a path integral in terms of an action projected to the pairs of
lattices $\{\Delta\Delta^*\}$.
The variables projected to the lattices are defined as follows.
{}For an edge $e$, define $U(e):=p\exp[\int_e dx^aA_a^i\tau_i]$, 
a path ordered parallel transport along $e$.
Here $\tau_i$ is the Pauli matrix divided by $2i$.
{}For a face $f$, define 
$\eta^i(f):={1\over2}\int_fdx^a dx^b B_{ab}^i$.
The scalar fields $\phi_{ij}$ are defined at the centers of faces.
The integrals over edge and face are oriented such that they are compatible 
with the orientation of $\Delta_p$ and $\Delta_p^*$.

The action projected to the pairs of lattices $\{\Delta\Delta^*\}$ is
defined by
\begin{eqnarray}
&S_{\{\Delta\Delta^*\}}[U,\eta,\phi]:=&
\sum_{p=1}^{P}\sum_{f\in\Delta_p\Delta_p^*}
\left( \eta^i(f)
{1\over2}{\rm Tr}[\tau_iU(f^*)\prod_{e\in f^*}^{path}U(e){\ }U^{-1}(f^*)]
\right.\nonumber\\&&\left.
-{\lambda\over2}\phi_{ij}(ff^*)
[\eta^i(f)\eta^j(f^*)+\eta^j(f)\eta^i(f^*)]
\right),
\label{eq:projection}
\end{eqnarray}
where $f^*$ is the face sharing its center with $f$ on a pair of
lattices $\Delta_p$ and $\Delta_p^*$,
and $ff^*$ means the center of $f$ and $f^*$.
$U(f^*)$ is a parallel transport defined not on an edge of the lattice 
but on a curve from the center of $f^*$ to one of the vertices 
belonging to $f^*$.
The choice of the vertex is arbitrary and does not affect the definition 
of the partition function of the model.
This parellel transport is necessary in order to make the first term 
gauge invariant since the two faces $f$ and $f^*$ intersect with each other
only at their center and the untraced parallel transport must base
on the center.

The exponential of the action can be considered as (an extension of)
the graph-cylindrical function discussed in \cite{AL}.
The integral measure we use for the variable $A_a^i$ is
the Ashtekar-Lewandowski measure \cite{AL}.
The measure for the scalar fields $\phi_{ij}$ is the one discussed 
in \cite{CR}.
The measure for the variable $B_{ab}^i$ is analogously defined.
The path integral has the form of
\begin{eqnarray}
&&
\int d\mu(A)d\mu(B)d\mu(\phi)\Psi_{\{\Delta\Delta^*\},\psi}(A,B,\phi):=
\nonumber\\&&
\int \prod_{e\in\{\Delta\Delta^*\}}dU(e) 
\prod_{f\in\{\Delta\Delta^*\}}d\eta(f)
\prod_{f\in\{\Delta\}}d\phi(ff^*)
\times\nonumber\\&&
\psi(U(e_1,A),\cdots U(e_l,A);\eta(f_1,B),\cdots \eta(f_m,B);
\phi(v_1),\cdots \phi(v_n)),
\end{eqnarray}
where $\Psi_{\{\Delta\Delta^*\},\psi}$ is a cylindrical function defined on
the pairs of the lattices $\{\Delta\Delta^*\}$ in terms of 
a complex valued integrable function 
$\psi$ on $[SU(2)]^l\times[su(2)]^m\times R^n$.
Note that $e\in\{\Delta\Delta^*\}$ does not mean that $e$ is a lattice
but means that $e$ is an edge of one of the lattices.
In the same way, $f\in\{\Delta\Delta^*\}$ means that 
$f$ is a face of one of the lattices.
We loosely use this kind of notation.
This integral form is diffeomorphism invariant.

The path integral which will be identified to (\ref{eq:partition}) is
\begin{eqnarray}
&&
{Z'}_{\beta,\lambda}[\xi]:=
\int dU\int_{-\infty}^{\infty}d\eta d\phi 
e^{i\beta\left[S_{\{\Delta\Delta^*\}}[U,\eta,\phi]
+\sum_{p=1}^{P}\sum_{f\in\Delta_p}\xi(ff^*)\eta_i(f)\eta^i(f^*)\right]}
\nonumber\\&&
=\prod_{e\in\{\Delta\Delta^*\}}\int dU(e)
\times
\prod_{f\in\{\Delta\Delta^*\}}\int_{-\infty}^{\infty}d\eta(f)
\times\nonumber\\&&
\prod_{f\in\{\Delta\}}\int_{-\infty}^{\infty}d\phi(ff^*)
\ \delta(\phi_{11}+\phi_{22}+\phi_{33})
\times\nonumber\\&&
\prod_{p=1}^{P}\prod_{f\in\Delta_p\Delta_p^*}
\int dU(f^*)e^{i\beta\eta^i(f){1\over2}
{\rm Tr}[\tau_iU(f^*)\prod_{e\in f^*}^{path}U(e)U^{-1}(f^*)]}
\times\nonumber\\&&
\prod_{p=1}^{P}\prod_{f\in\Delta_p}
e^{-{i\over2}\beta\lambda\phi_{ij}(ff^*)
[\eta^i(f)\eta^j(f^*)+\eta^j(f)\eta^i(f^*)]}
e^{i\beta\xi(ff^*)\eta_i(f)\eta^i(f^*)}.
\label{eq:pathintegral}
\end{eqnarray}
Here, the external field $\xi$ is defined at the centers of faces
as $\phi_{ij}$ is defined.
We note that if $P=1$ then
one can easily solve the constraints on
$\eta^i$ imposed by $\phi_{ij}$ and finds only degenerate solutions 
corresponding to $\delta_{ij}\tilde\epsilon^{abcd}B_{ab}^iB_{cd}^j=0$.
In order to include the non-degenerate sector, one needs $P>1$.
(See Appendix \ref{app:non-deg} for more discussion on 
the non-degenerate sector.)

Let us
expand the exponential containing the trace of an SU(2) element
to its characters as follows.
\begin{eqnarray}
&& 
\int dV
e^{i\beta\eta^i{1\over2}{\rm Tr}[\tau_iVUV^{-1}]}=
\int dV
\sum_{l_1,l_2,l_3}{J_{2l_1+1}(\beta\eta^1)\over\beta\eta^1}
{J_{2l_2+1}(\beta\eta^2)\over\beta\eta^2}
{J_{2l_3+1}(\beta\eta^3)\over\beta\eta^3}
\times\nonumber\\&&
2^3(-1)^{l_1+l_2+l_3}
(2l_1+1)(2l_2+1)(2l_3+1)
\times\nonumber\\&&
\chi_{l_1}[\tau_1VUV^{-1}]\chi_{l_2}[\tau_2VUV^{-1}]
\chi_{l_3}[\tau_3VUV^{-1}]
\nonumber\\&&
=\sum_{l_1,l_2,l_3}{J_{2l_1+1}(\beta\eta^1)\over\beta\eta^1}
{J_{2l_2+1}(\beta\eta^2)\over\beta\eta^2}
{J_{2l_3+1}(\beta\eta^3)\over\beta\eta^3}
\sum_{j}\Omega^{(j)}(l_1,l_2,l_3)\chi_j(U).
\end{eqnarray}
Here, $J_m(x)$ is the Bessel function of the first kind and we have
used the following formulae.
\begin{eqnarray}
&&
e^{ix{1\over2}{\rm Tr}U}=\sum_j 2{2j+1\over x}i^{2j}J_{2j+1}(x)\chi_j(U),
\\&&
J_m(x)={1\over2\pi}\int_{-\pi}^{\pi}d\theta e^{ix\sin\theta-im\theta}.
\end{eqnarray}
The first formula can be easily proved and the second is a definition 
of the Bessel function.
$\Omega^{(j)}(l_1,l_2,l_3)$ is defined as follows.
\begin{eqnarray}
&&
\Omega^{(j)}(l_1,l_2,l_3):=2^3(-1)^{l_1+l_2+l_3}
(2l_1+1)(2l_2+1)(2l_3+1)
\times\nonumber\\&&
\sum_{m_i,n_i,m,n}D_{mn}^{(j)}(1)
D_{m_1n_1}^{(l_1)}(\tau_1)D_{m_2n_2}^{(l_2)}(\tau_2)
D_{m_3n_3}^{(l_3)}(\tau_3)(-1)^{n_3-m_3}
\sum_{l_{12},m_{12},n_{12}}(2l_{12}+1)
\times\nonumber\\&&
\left(\matrix{l_1&l_2&l_{12}\cr n_1&n_2&n_{12}}\right)
\left(\matrix{l_1&l_2&l_{12}\cr m_1&m_2&m_{12}}\right)
\left(\matrix{l_{12}&l_3&j\cr -n_{12}&n_3&n}\right)
\left(\matrix{l_{12}&l_3&j\cr -m_{12}&m_3&m}\right).
\label{eq:preOmega}
\end{eqnarray}
Here, $D_{mn}^{(j)}(U)$ is the spin-j representation matrix of SU(2) 
element $U$
and $m$ and $n$ run from $-j$ through $j$ with the increment 1.
$\left(\matrix{j_1&j_2&j\cr m_1&m_2&m}\right)$ 
is the so-called 3j-coefficient and 
defined by 
\begin{eqnarray}
&&
\left(\matrix{j_1&j_2&j\cr m_1&m_2&m}\right):=
{(-1)^{j_1-j_2-m}\over\sqrt{2j+1}}
\langle j_1m_1;j_2m_2|j,-m\rangle,
\end{eqnarray}
with the Clebsch-Gordan coefficient $\langle j_1m_1;j_2m_2|jm\rangle$
and we have used properties of $D_{nm}^{(j)}(U)$ such as
\begin{eqnarray}
&&(-1)^{n-m}D_{nm}^{(j)}(U)=D_{-n,-m}^{(j)*}(U),
\\&&
D_{n_1m_1}^{(j_1)}(U)D_{n_2m_2}^{(j_2)}(U)=
\sum_{j,n,m}(2j+1)
\left(\matrix{j_1&j_2&j\cr n_1&n_2&n}\right)
\left(\matrix{j_1&j_2&j\cr m_1&m_2&m}\right)
D_{nm}^{(j)*}(U),{\ \ }
\\&&
\int dU D_{mn}^{(i)}(U)D_{m'n'}^{(j)*}(U)=
{1\over 2j+1}\delta_{ij}\delta_{mm'}\delta_{nn'},
\end{eqnarray}
where the asterisks mean the complex conjugate and the sum of $j$ 
is taken over $|j_1-j_2|$ through $j_1+j_2$ and the sums of $n$ and $m$
over $-j$ through $j$.
Then the partition function can be written as follows.
\begin{eqnarray}
&&
{Z'}_{\beta,\lambda}[\xi]=
\prod_{e\in\{\Delta\Delta^*\}}\int dU(e)
\times
\prod_{f\in\{\Delta\Delta^*\}}\int_{-\infty}^{\infty}d\eta(f)
\times\nonumber\\&&
\prod_{f\in\{\Delta\}}\int_{-\infty}^{\infty}d\phi(ff^*)
\ \delta(\phi_{11}+\phi_{22}+\phi_{33})
\times\nonumber\\&&
\prod_{p=1}^{P}\prod_{f\in\Delta_p}\sum_{\{l\}_{ff*}}
{J_{2l_1(f^*)+1}(\beta\eta^1(f))\over\beta\eta^1(f)}
{J_{2l_2(f^*)+1}(\beta\eta^2(f))\over\beta\eta^2(f)}
{J_{2l_3(f^*)+1}(\beta\eta^3(f))\over\beta\eta^3(f)}
\times\nonumber\\&&
{J_{2l_1(f)+1}(\beta\eta^1(f^*))\over\beta\eta^1(f^*)}
{J_{2l_2(f)+1}(\beta\eta^2(f^*))\over\beta\eta^2(f^*)}
{J_{2l_3(f)+1}(\beta\eta^3(f^*))\over\beta\eta^3(f^*)}
\times\nonumber\\&&
e^{-{i\over2}\beta\lambda\phi_{ij}(ff^*)
[\eta^i(f)\eta^j(f^*)+\eta^j(f)\eta^i(f^*)]}
e^{i\beta\xi(ff^*)\eta_i(f)\eta^i(f^*)}
\times\nonumber\\&&
\sum_{j(f)}\Omega^{(j(f))}(l_1(f),l_2(f),l_3(f))
\chi_{j(f)}(\prod_{e\in f}^{path}U(e))
\times\nonumber\\&&
\sum_{j(f^*)}
\Omega^{(j(f^*))}(l_1(f^*),l_2(f^*),l_3(f^*))
\chi_{j(f^*)}(\prod_{e\in f^*}^{path}U(e)).
\end{eqnarray}
Here, $\{l\}_{ff*}$ stands for 
$l_1(f),l_2(f),l_3(f),l_1(f^*),l_2(f^*),l_3(f^*)$.
Let us write the other exponentials explicitly and 
expand them as a power series as follows.
\begin{eqnarray}
&&
e^{-{i\over2}\beta\lambda\phi_{ij}(ff^*)
[\eta^i(f)\eta^j(f^*)+\eta^j(f)\eta^i(f^*)]}
e^{i\beta\xi(ff^*)\eta_i(f)\eta^i(f^*)}
\nonumber\\&&
=e^{-i\beta\lambda\phi_{11}(ff^*)\eta^1(f)\eta^1(f^*)}
e^{-i\beta\lambda\phi_{22}(ff^*)\eta^2(f)\eta^2(f^*)}
\times\nonumber\\&&
e^{-i\beta\lambda\phi_{33}(ff^*)\eta^3(f)\eta^3(f^*)}
e^{-i\beta\lambda\phi_{12}(ff^*)
[\eta^1(f)\eta^2(f^*)+\eta^2(f)\eta^1(f^*)]}
\times\nonumber\\&&
e^{-i\beta\lambda\phi_{23}(ff^*)
[\eta^2(f)\eta^3(f^*)+\eta^3(f)\eta^2(f^*)]}
e^{-i\beta\lambda\phi_{31}(ff^*)
[\eta^3(f)\eta^1(f^*)+\eta^1(f)\eta^3(f^*)]}
\times\nonumber\\&&
e^{i\beta\xi(ff^*)
[\eta^1(f)\eta^1(f^*)+\eta^2(f)\eta^2(f^*)+\eta^3(f)\eta^3(f^*)]}
\nonumber\\&&
=\sum_{N_{11}=0}^{\infty}\sum_{M_{1}=0}^{\infty}
{[-i\beta\lambda\phi_{11}(ff^*)]^{N_{11}}\over N_{11}!}
{[i\beta\xi(ff^*)]^{M_{1}}\over M_{1}!}
[\eta^1(f)\eta^1(f^*)]^{N_{11}+M_{1}}
\times\nonumber\\&&
\sum_{N_{22}=0}^{\infty}\sum_{M_{2}=0}^{\infty}
{[-i\beta\lambda\phi_{22}(ff^*)]^{N_{22}}\over N_{22}!}
{[i\beta\xi(ff^*)]^{M_{2}}\over M_{2}!}
[\eta^2(f)\eta^2(f^*)]^{N_{22}+M_{2}}
\times\nonumber\\&&
\sum_{N_{33}=0}^{\infty}\sum_{M_{3}=0}^{\infty}
{[-i\beta\lambda\phi_{33}(ff^*)]^{N_{33}}\over N_{33}!}
{[i\beta\xi(ff^*)]^{M_{3}}\over M_{3}!}
[\eta^3(f)\eta^3(f^*)]^{N_{33}+M_{3}}
\times\nonumber\\&&
\sum_{N_{12}=0}^{\infty}\sum_{N_{21}=0}^{\infty}
{[-i\beta\lambda\phi_{12}(ff^*)]^{N_{12}+N_{21}}
\over N_{12}!N_{21}!}
[\eta^1(f)\eta^2(f^*)]^{N_{12}}[\eta^2(f)\eta^1(f^*)]^{N_{21}}
\times\nonumber\\&&
\sum_{N_{23}=0}^{\infty}\sum_{N_{32}=0}^{\infty}
{[-i\beta\lambda\phi_{23}(ff^*)]^{N_{23}+N_{32}}
\over N_{23}!N_{32}!}
[\eta^2(f)\eta^3(f^*)]^{N_{23}}[\eta^3(f)\eta^2(f^*)]^{N_{32}}
\times\nonumber\\&&
\sum_{N_{31}=0}^{\infty}\sum_{N_{13}=0}^{\infty}
{[-i\beta\lambda\phi_{31}(ff^*)]^{N_{31}+N_{13}}
\over N_{31}!N_{13}!}
[\eta^3(f)\eta^1(f^*)]^{N_{31}}[\eta^1(f)\eta^3(f^*)]^{N_{13}}.
\end{eqnarray}
Here, the non-negative integers $N_{ij}$ and $M_{i}$ with $i,j=1,2,3$ 
are associated with
the pair of faces $f$ and $f^*$.
Then the partition function becomes
\begin{eqnarray}
&&
{Z'}_{\beta,\lambda}[\xi]
=\prod_{e\in\{\Delta\Delta^*\}}\int dU(e)\times
\prod_{p=1}^{P}\prod_{f\in\Delta_p}\sum_{\{l\}_{ff^*}}
W_{\beta,\lambda}(\{l\}_{ff^*},\xi)
\times\nonumber\\&&
\sum_{j(f)}\Omega^{(j(f))}(l_1(f),l_2(f),l_3(f))
\chi_{j(f)}(\prod_{e\in f}^{path}U(e))
\times\nonumber\\&&
\sum_{j(f^*)}
\Omega^{(j(f^*))}(l_1(f^*),l_2(f^*),l_3(f^*))
\chi_{j(f^*)}(\prod_{e\in f^*}^{path}U(e)),
\end{eqnarray}
with
\begin{eqnarray}
&&
W_{\beta,\lambda}(\{l\}_{ff^*},\xi):=
\int_{-\infty}^{\infty}d\eta(f) d\eta(f^*)
\int_{-\infty}^{\infty}d\phi(ff^*)
\ \delta(\phi_{11}+\phi_{22}+\phi_{33})
\times\nonumber\\&&
\sum_{\{N,M\}_{ff^*}}
{[-i\beta\lambda\phi_{11}(ff^*)]^{N_{11}}\over N_{11}!}
{[-i\beta\lambda\phi_{22}(ff^*)]^{N_{22}}\over N_{22}!}
{[-i\beta\lambda\phi_{33}(ff^*)]^{N_{33}}\over N_{33}!}
\times\nonumber\\&&
{[-i\beta\lambda\phi_{12}(ff^*)]^{N_{12}+N_{21}}
\over N_{12}!N_{21}!}
{[-i\beta\lambda\phi_{23}(ff^*)]^{N_{23}+N_{32}}
\over N_{23}!N_{32}!}
{[-i\beta\lambda\phi_{31}(ff^*)]^{N_{31}+N_{13}}
\over N_{31}!N_{13}!}
\times\nonumber\\&&
{[i\beta\xi(ff^*)]^{M_{1}+M_{2}+M_{3}}
\over M_{1}!M_{2}!M_{3}!}
\times\nonumber\\&&
[\eta^1(f)]^{M_{1}+N_{11}+N_{12}+N_{13}}
[\eta^2(f)]^{M_{2}+N_{22}+N_{21}+N_{23}}
[\eta^3(f)]^{M_{3}+N_{33}+N_{32}+N_{31}}
\times\nonumber\\&&
[\eta^1(f^*)]^{M_{1}+N_{11}+N_{21}+N_{31}}
[\eta^2(f^*)]^{M_{2}+N_{22}+N_{12}+N_{32}}
[\eta^3(f^*)]^{M_{3}+N_{33}+N_{23}+N_{13}}
\times\nonumber\\&&
{J_{2l_1(f^*)+1}(\beta\eta^1(f))\over\beta\eta^1(f)}
{J_{2l_2(f^*)+1}(\beta\eta^2(f))\over\beta\eta^2(f)}
{J_{2l_3(f^*)+1}(\beta\eta^3(f))\over\beta\eta^3(f)}
\times\nonumber\\&&
{J_{2l_1(f)+1}(\beta\eta^1(f^*))\over\beta\eta^1(f^*)}
{J_{2l_2(f)+1}(\beta\eta^2(f^*))\over\beta\eta^2(f^*)}
{J_{2l_3(f)+1}(\beta\eta^3(f^*))\over\beta\eta^3(f^*)}.
\end{eqnarray}
Here, $\{N\}_{ff^*}$ stands for the non-negative integers $N_{ij}$
and $M_{i}$ 
associated with the pair of faces $f$ and $f^*$.
The integrations over $\eta^i$ can be performed.
Here, for concreteness, we compute the coefficient for the case that 
$\beta$ is positive real number.  
A useful formula is
$J(m,n):=\int_{-\infty}^{\infty}x^{n-1}J_m(x)dx=
\beta^n\int_{-\infty}^{\infty}x^{n-1}J_m(\beta x)dx$ and
\begin{eqnarray}
&& J(m,n\ge 1)=i^{n-1}
\left[\left({d\over d\theta}{1\over\cos\theta}\right)^{n-1}e^{-im\theta}
|_{\theta=0}
+\left({d\over d\theta}{1\over\cos\theta}\right)^{n-1}e^{-im\theta}
|_{\theta=\pi}
\right],
\\&&
J(m,0)={1\over m}[1-(-1)^m].
\end{eqnarray}
Here $\left({d\over d\theta}{1\over\cos\theta}\right)^{n-1}$ means
${d\over d\theta}{1\over\cos\theta}$ acts $n-1$ times.
For given $m$ and $n$, $J(m,n)$ exists and can be explicitly computed.
In terms of $J(m,n)$, the coefficient $W_{\beta,\lambda}(\{l\}_{ff^*},\xi)$
can be written as follows.
\begin{eqnarray}
&&
W_{\beta,\lambda}(\{l\}_{ff^*},\xi)=\beta^{-6}
\int_{-\infty}^{\infty}d\phi(ff^*)
\ \delta(\phi_{11}+\phi_{22}+\phi_{33})
\times\nonumber\\&&
\sum_{\{N,M\}_{ff^*}}
{[-i\beta^{-1}\lambda\phi_{11}(ff^*)]^{N_{11}} \over N_{11}!}
{[-i\beta^{-1}\lambda\phi_{22}(ff^*)]^{N_{22}} \over N_{22}!}
{[-i\beta^{-1}\lambda\phi_{33}(ff^*)]^{N_{33}} \over N_{33}!}
\times\nonumber\\&&
{[-i\beta^{-1}\lambda\phi_{12}(ff^*)]^{N_{12}+N_{21}}
\over N_{12}!N_{21}!}
{[-i\beta^{-1}\lambda\phi_{23}(ff^*)]^{N_{23}+N_{32}}
\over N_{23}!N_{32}!}
{[-i\beta^{-1}\lambda\phi_{31}(ff^*)]^{N_{31}+N_{13}}
\over N_{31}!N_{13}!}
\times\nonumber\\&&
{[i\beta^{-1}\xi(ff^*)]^{M_{1}+M_{2}+M_{3}}
\over M_{1}!M_{2}!M_{3}!}
\times\nonumber\\&&
J(2l_1(f^*)+1,M_{1}+N_{11}+N_{12}+N_{13})
J(2l_2(f^*)+1,M_{2}+N_{22}+N_{21}+N_{23})
\times\nonumber\\&&
J(2l_3(f^*)+1,M_{3}+N_{33}+N_{32}+N_{31})
J(2l_1(f)+1,M_{1}+N_{11}+N_{21}+N_{31})
\times\nonumber\\&&
J(2l_2(f)+1,M_{2}+N_{22}+N_{12}+N_{32})
J(2l_3(f)+1,M_{3}+N_{33}+N_{23}+N_{13}).
\label{eq:preW}
\end{eqnarray}

Now, define the coefficients $\Omega^{(j)}(\{l\})$ and
$W_{\beta,\lambda}(\{l\},\xi)$ such that
${Z'}_{\beta,\lambda}[\xi]$ is identified to $Z_{\beta,\lambda}[\xi]$
as follows.
\begin{eqnarray}
&&\Omega^{(j)}(\{l\}):=
\prod_{p=1}^{P}
\prod_{f\in\Delta_p\Delta_p^*}2^{-9}
\Omega^{(j(f))}(l_1(f),l_2(f),l_3(f)),
\label{eq:Omega}
\\&&
W_{\beta,\lambda}(\{l\},\xi):=
\prod_{p=1}^{P}
\prod_{f\in\Delta_p}\beta^6
W_{\beta,\lambda}(\{l\}_{ff^*},\xi).
\label{eq:W}
\end{eqnarray}
Here, $2^{-9}$ and $\beta^6$ have been inserted for convenient 
normalization.
(\ref{eq:Omega}) with (\ref{eq:preOmega}), and (\ref{eq:W}) with
(\ref{eq:preW}) are the final forms for the coefficients
(\ref{eq:coefficients}) in the partition function of the model.

{}For a completeness, let us compute the coefficient 
$C_{\beta,\lambda}^{(j)}[\xi]$  for the case that $\beta=1$, $\lambda=0$
and $\xi=0$.
In that case,
\begin{eqnarray}
&&
W_{1,0}(\{l\},0)=\prod_{p=1}^{P}
\prod_{f\in\Delta_p\Delta_p^*}
J(2l_1(f)+1,0){\ }J(2l_2(f)+1,0){\ }J(2l_3(f)+1,0)
\nonumber\\&&
=\prod_{p=1}^{P}
\prod_{f\in\Delta_p\Delta_p^*}
\left[{1-(-1)^{2l_1(f)+1}\over 2l_1(f)+1}\right]
\left[{1-(-1)^{2l_2(f)+1}\over 2l_2(f)+1}\right]
\left[{1-(-1)^{2l_3(f)+1}\over 2l_3(f)+1}\right],
\end{eqnarray}
and
\begin{eqnarray}
&&
C_{1,0}^{(j)}[0]=\sum_{\{l\}}W_{1,0}(\{l\},0)\Omega^{(j)}(\{l\})
\nonumber\\&&
=\prod_{p=1}^{P}
\prod_{f\in\Delta_p\Delta_p^*}\sum_{l_1,l_2,l_3}
[1-(-1)^{2l_1+1}][1-(-1)^{2l_2+1}][1-(-1)^{2l_3+1}]2^{-6}
(-1)^{l_1+l_2+l_3}
\times\nonumber\\&&
\sum_{m_i,n_i,m,n}D_{mn}^{(j)}(1)D_{m_1n_1}^{(l_1)}(\tau_1)
D_{m_2n_2}^{(l_2)}(\tau_2)D_{m_3n_3}^{(l_3)}(\tau_3)
(-1)^{n_3-m_3}
\sum_{l_{12},m_{12},n_{12}}(2l_{12}+1)
\times\nonumber\\&&
\left(\matrix{l_1&l_2&l_{12}\cr n_1&n_2&n_{12}}\right)
\left(\matrix{l_1&l_2&l_{12}\cr m_1&m_2&m_{12}}\right)
\left(\matrix{l_{12}&l_3&j\cr -n_{12}&n_3&n}\right)
\left(\matrix{l_{12}&l_3&j\cr -m_{12}&m_3&m}\right)
\nonumber\\&&
=\prod_{p=1}^{P}
\prod_{f\in\Delta_p\Delta_p^*}\sum_{l_1,l_2,l_3,l_{12}}
\chi_{l_1}(\tau_1)\chi_{l_2}(\tau_2)
\chi_{l_3}(\tau_3)\chi_{l_{12}}(1)
\times\nonumber\\&&
\int dV dW \chi_{l_1}(\tau_1V)\chi_{l_2}(\tau_2V)
\chi_{l_{12}}(VW^{-1})\chi_{l_3}(\tau_3W)\chi_{j}(W).
\end{eqnarray}
Here, we have used the formulae that $\chi_l(\tau_i)={1\over2}(-1)^l$
if $l$ is an integer and $\chi_l(\tau_i)=0$ if $l$ is a half-integer
and that $\chi_l(1)=2l+1$ and that
\begin{eqnarray}
&&
\left(\matrix{j_1&j_2&j_3\cr n_1&n_2&n_3}\right)
\left(\matrix{j_1&j_2&j_3\cr m_1&m_2&m_3}\right)=
\int dV D_{n_1m_1}^{(j_1)}(V)D_{n_2m_2}^{(j_2)}(V)D_{n_3m_3}^{(j_3)}(V).
\end{eqnarray}
Let us work out the coefficient $C_{1,0}^{(j)}[0]$.
{}From the last expression, $V$ and $W$ are forced to be $V=W=1$ after 
the sums over $l_1$, $l_2$, $l_3$ and $l_{12}$ and the integrations. 
Then we get
\begin{eqnarray}
&&
C_{1,0}^{(j)}[0]=\prod_{p=1}^{P}
\prod_{f\in\Delta_p\Delta_p^*}\chi_j(1)=
\prod_{p=1}^{P}
\prod_{f\in\Delta_p\Delta_p^*}(2j+1)
\end{eqnarray}
if $j$ is an integer; otherwise, vanishes.
Here, $j$ is associated with each $f$.
This ``trivial" coefficient makes the partition function of the model
the multiple power of that of the lattice BF theory as claimed already.

{}Finally, for a physical application, we present a computation formula
for spacetime volume density correlations.
{}From the partition function of the model, define
$Z_{\beta,\lambda}^{(x_1\cdots x_n)}[\xi]$ which is
$Z_{\beta,\lambda}[\xi]$ with $W_{\beta,\lambda}(\{l\},\xi)$ replaced by
${\delta\over\delta\xi(x_1)}\cdots{\delta\over\delta\xi(x_n)}  
W_{\beta,\lambda}(\{l\},\xi)$.
Then the spacetime volume density correlation function is
\begin{eqnarray}
&&\langle\sqrt{g(x_1)}\cdots\sqrt{g(x_n)}\rangle:=
{Z_{\beta,\lambda}^{(x_1\cdots x_n)}[0]\over Z_{\beta,\lambda}[0]}.
\end{eqnarray}
This quantity is the average of a product of spacetime volume densities 
defined at the centers of faces with respect to the partition function.
The spacetime volume density at $x$ is computed by
\begin{eqnarray}
&&
\sqrt{g(x)}:=\sum_{ff^*=x}\eta_i(f)\eta^i(f^*),
\end{eqnarray}
where the sum is taken over faces $f$ and $f^*$ whose center $ff^*$
is located at $x$.


\section{Conclusion}\label{sec:conclusion}

We have presented a model for quantum gravity, 
which could reveal an aspect of the spin-foam interpretation of spacetime.
Its partition function is defined by (\ref{eq:partition}) 
with coefficients (\ref{eq:coefficients}).
We have determined the coefficients based on a path integral of
(3+1) dimensional general relativity in Riemannian spacetime.
To do so, we projected the Plebanski action (\ref{eq:action}) 
to multiple pairs of lattices, 
each pair of which consists of two lattices dual to each other, 
(\ref{eq:projection})
and then computed its path integral (\ref{eq:pathintegral}).
The final forms of the coefficients are given by
(\ref{eq:Omega}) with (\ref{eq:preOmega}), 
and (\ref{eq:W}) with (\ref{eq:preW}).
In terms of the partition function, we have presented a computation
formula for spacetime volume density correlations.

We conclude with remarks for further investigations:

(i)
The variables $U$ and $\phi_{ij}$ are left unintegrated.
The integration for $U$ is known to produce functions of spins and 
intertwiners labeling respectively the faces  and edges of a spin-foam.
This integration is already present in the case of the BF theory and 
the resulting functions consist of the so-called 3nj-symboles.
A spin-foam can be decomposed to a family of surfaces and 
these functions can be understood as telling the number of 
``distinguishable" ways with respect to a given lattice 
(dual to the lattice the spin-form bases on)
the family of surfaces are allocated as studied in \cite{JI}.
The integrations for $\phi_{ij}$ are particular in quantization of 
general relativity and are supposed to induce local degrees of freedom.
In the present model, physics of the local degrees of freedom
may be translated from the information how single spin-foams 
``interact" with each other.
The coefficient $W_{\beta,\lambda}(\{l\},\xi)$
has roughly the form of
$\int dx_1\cdots dx_k \sum_{n_1\cdots n_k}f_{n_1\cdots n_k}
{x_1^{n_1}\cdots x_k^{n_k}\over n_1!\cdots n_k!}$ with $x's$ 
corresponding to
$\phi_{ij}$ and naive term-by-term integration leads to divergences.
We have to understand the properties of $f_{n_1\cdots n_k}$ before 
the integrations of $\phi_{ij}$.

(ii)
To perform the path integral of general relativity, we have  
used diffeomorphism invariant measures.
However, we have to check the resulting partition function is 
really diffeomorphism invariant.
In a limiting case, we have shown that the partition function 
is reduced to
the multiple power of the partition function of the lattice BF theory,
which is a topological invariant and certainly diffeomorphism invariant.

(iii)
A technical peculiarity in the construction of the present model
is the use of multiple pairs of lattices. 
Each pair consists of two lattices dual to each other.
Because of this, physics comes out not from single spin-foams
but from ``interactions" between them.
We would like to understand whether the use of the multiple lattices 
is necessary or redundant.

(iv)
In the standard lattice gauge theory, 
in which the number of degrees of freedom is
finite and each variable is compactified,
the use of lattice reguralizes the divergences due to the integrals along
the gauge orbits as long as computations of gauge invariant quantities 
are concerned.
However, in the present model, the reguralization is not yet under control
because not all the variables are compactified.
In particular, to perform computations of gauge dependent quantities 
such as propagaters, the knowledge of how to fix a gauge is crucial. 

(v)
As physical applications, one wants to compute physical quantities. 
{}For example, a detailed study of
the spacetime volume density correlation function we have presented
is to be pursued.

(vi)
In order to understand the physical meaning of spin-foams, 
a possible way of studying it is the following.
Couple external fields to the variables $A_a^i$, $B_{ab}^i$ and $\phi_{ij}$ 
and then compute the partition function with the external field.
In terms of the partition function, one may define Schwinger equations
corresponding to the classical Einstein equations with those variables.
Detailed study may allow one to understand the physical meaning of
the spin-foams corresponding to classical spacetime geometries.

(vii)
Ultimately one wants to formulate a quantum gravity model in 
Lorentizn spacetime.
As an extension of the present model, 
Lorentzian spacetime gravity is defined in terms of
the path integral of complex variables with certain reality conditions.
A possibility of including the reality conditions into the path integral
ought to be studied.


\appendix

\section{Pairs of lattices}\label{app:pairs}

We discuss in this appendix a way of constructing the multiple pairs of
lattices described in Sec.\ref{sec:model}.
Because 4-dimensional lattices are difficult to visualize, we first utilize 
2-dimensional lattices , the role of whose edges and dual edges mimics the
role of faces and dual faces of 4-dimensional lattices, to construct
pairs of lattices.
Then we repeat the construction in 4-dimensional case.
Note that these constructions are not oriented to show a mathematical
proof of the existence of the pairs of lattices.

On a 2-dimensional manifold with the topology of $S^2$ for simplicity, 
draw a lattice and its dual lattice.
They consist of 0-cells (vertices), 1-cells (edges), and 2-cells 
(polygons or faces).
The edges are not necessarily straight lines but may be curved.
Every edge of the lattice (and the dual lattice) intersects an edge of
the dual lattice (and the lattice respectively) at their ``center.''
Let us plot a dot at every center of edges (and dual edges).
Draw curves connecting the every two centers of edges sharing a vertx of
the lattices.
This drawing creates another lattice, whose vertices are located at
the centers of the edges of the original lattices.
We call the created lattice ``child'' of the original lattices.
Erase the original lattices.
We have now only the child on the 2-dimensional manifold with dots indicating
the positions of the vertices of the child.

Move the child as follows while the dots are fixed.
Every vertex is moved from one dot to another or is not moved 
so that every dot is occupied by only one vertex.
The edges sharing a vertex are moved together with the vertex if the vertex 
is moved.
Let us call the resulting lattice ``cousin'' of the child.
They are isomorphic to each other.

Draw a pair of lattices (a lattice and its dual lattice) from the cousin
such that their relationship is analogous to the relationship between the 
original pair of lattices and their child.
Erase the cousin.
Now, we have only the second pair of lattices with dots on the 2-dimensional
manifold.
The centers of edges of this pair of lattices are located at the dots
and hence the position of the center of every edge of this pair of lattices
coincides with the position of the center of an edge of the original pair
of lattices.
One of the original pair is isomorphic to one of the second pair
while the other lattice of the original pair is isomorphic to
the other lattice of the second pair.
By repeating the same process one can generate a third pair, a fourth pair,
and so forth.

Next, let us repeat the analogous construction for 4-dimensional case.
In a 4-dimensional spacetime manifold with the topology of $S^4$
for simplicity,
draw a lattice and its dual lattice.
They consist of 0-cells (vertices), 1-cells (edges), 2-cells 
(polygons or faces), 3-cells (polyhedrons) and 4-cells.
The edges and polygons are not necessarily straight lines or flat surfaces
respectively but may be curved ones. 
Every face of the lattice (and the dual lattice) intersects a face
of the dual lattice (and the lattice respectively) at their ``center.''
Let us plot a dot at every center of faces (and dual faces).
Draw curves connecting the every two centers of faces sharing a vertex of
the lattices.
The curves connecting the centers of faces sharing a vertex form a polygon.
This drawing creates another lattice.
Its vertices are located at the centers of the faces of
the original lattices.
We call the created lattice ``child'' of the original lattices.
Erase the original lattices.
We have now only the child in the spacetime manifold with dots
indicating the positions of the vertices of the child.

Move the child as follows while the dots are fixed.
Every vertex is moved from one dot to another or is not moved 
so that every dot is occupied by only one vertex.
The edges sharing a vertex are moved together with the vertex if the vertex 
is moved.
Let us call the resulting lattice ``cousin'' of the child.
They are isomorphic to each other.

Draw a pair of lattices (a lattice and its dual lattice) from the cousin
such that their relationship is analogous to the relationship between the 
original pair of lattices and their child.
Erase the cousin.
Now, we have only the second pair of lattices with dots in the spacetime
manifold.
The centers of faces of this pair of lattices are located at the dots
and hence the position of the center of every face of this pair of lattices
coincides with the position of the center of a face of the original pair
of lattices.
One of the original pair is isomorphic to one of the second pair
while the other lattice of the original pair is isomorphic to
the other lattice of the second pair.
By repeating the same process one can generate a third pair, a fourth pair,
and so forth.

\section{Non-degenerate sector}\label{app:non-deg}

Pairs of lattices in 4-dimensions are used in the model.
Each pair consists of two lattices dual to each other.
That is, every k-cell of one lattice intersects with a (4-k)-cell of
the other lattice at a point inside the cells.
The intersection is called ``center.''
In particular, every face (2-cell) of one lattice shares a center with
a face of the other lattice.
These two faces are said to be dual to each other.
Let $\Delta_p$ and $\Delta_p^*$ denote a pair of lattices with
$p=1,\cdots,P$.
The $P$ is the number of pairs.
If they form the multiple pairs described in Sec.\ref{sec:model}, then
there exist a face $f_p$ in $\Delta_p$ and its dual face $f_p^*$ 
in $\Delta_p^*$ with their center
denoted by $f_pf_p^*$ such that $f_1f_1^*=\cdots= f_Pf_P^*$.
(Here the equality means that their positions coincide.)

Let $ff^*$ represent the position of the coincident centers.
The Lagrange multipliers $\phi_{ij}$ are defined at $ff^*$.
The constraints imposed by the multipliers at $ff^*$ are
\begin{eqnarray}
&&
\sum_{p=1}^P[\eta^i(f_p)\eta^j(f_p^*)+\eta^j(f_p)\eta^i(f_p^*)]
=C(ff^*)\delta^{ij}.
\end{eqnarray}
Here $C(ff^*)$ is an arbitrary non-negative real number determined by
$\eta^i(f_p)$ and $\eta^i(f_p^*)$ and proportional to spacetime
volume density at $ff^*$.
$C(ff^*)>0$ and $C(ff^*)=0$ mean that 
the solutions for $\eta^i$ are non-degenerate and degenerate 
respectively at $ff^*$.

It is easy to check that if $P=1$, then solutions for $\eta^i$ at $ff^*$
do not exist unless $C(ff^*)=0$.
This fact means that the model needs more than a pair of lattices
in order to contain the non-degenerate sector.
Once the number of pairs, $P$, is fixed, then the non-degenerate and 
degenerate sectors can be clearly characterized.
Since the constraints have the form of the inner product for three
vectors ($i,j=1,2,3$) with their norms proportional to spacetime
volume density, the non-degenerate sector of the model is 
analogous to the space spanned by non-degenerate orthogonal vectors
in a 3-dimensional linear space.


\end{document}